\documentclass[acmsmall,screen]{acmart}
\AtBeginDocument{%
  }

\setcopyright{cc}
\setcctype{by}
\acmDOI{10.1145/3715742}
\acmYear{2025}
\acmJournal{PACMSE}
\acmVolume{2}
\acmNumber{FSE}
\acmArticle{FSE027}
\acmMonth{7}
\received{2024-09-13}
\received[accepted]{2025-01-14}

\usepackage{amsmath}
\usepackage{caption} 

\begin{document}

\title{Cross-System Categorization of Abnormal Traces in Microservice-Based Systems via Meta-Learning}


\author{Yuqing Wang}
\orcid{0000-0003-0175-005X}
\authornote{Yuqing Wang is the corresponding author (yuqing.wang@helsinki.fi) }
\email{yuqing.wang@helsinki.fi}
\authornotemark[0]
\affiliation{%
  \institution{University of Helsinki}
  \city{Helsinki}
  \country{Finland}
}


\author{Mika V. M{\"a}ntyl{\"a}}
\orcid{0000-0002-2841-5879}
\affiliation{%
  \institution{University of Helsinki}
  \city{Helsinki}
  \country{Finland}}

\affiliation{%
  \institution{University of Oulu}
  \city{Oulu}
  \country{Finland}}
\email{mika.mantyla@helsinki.fi} 

\author{Serge Demeyer}
\orcid{0000-0002-4463-2945}
\affiliation{%
  \institution{University of Antwerp}
  \country{Belgium}}
\email{serge.demeyer@uantwerpen.be}

\author{Mutlu Beyaz{\i}t}
\orcid{0000-0003-2714-8155}
\affiliation{%
  \institution{University of Antwerp}
  \country{Belgium}}
\email{mutlu.beyazit@uantwerpen.be}

\author{Joanna Kisaakye}
\orcid{0000-0001-7081-5385}
\affiliation{%
  \institution{University of Antwerp}
  \country{Belgium}}
\email{joanna.kisaakye@uantwerpen.be}

\author{Jesse Nyss{\"o}l{\"a}}
\orcid{0009-0006-7276-5696}
\affiliation{%
  \institution{University of Helsinki}
  \country{Finland}}
\email{jesse.nyyssola@helsinki.fi}

\renewcommand{\shortauthors}{Wang et al.}

\begin{abstract}
Microservice-based systems (MSS) may fail with various fault types, due to their complex and dynamic nature. While existing AIOps methods excel at detecting abnormal traces and locating the responsible service(s), human efforts from practitioners are still required for further root cause analysis to diagnose specific fault types and analyze failure reasons for detected abnormal traces, particularly when abnormal traces do not stem directly from specific services. In this paper, we propose a novel AIOps framework, TraFaultDia, to automatically classify abnormal traces into fault categories for MSS. We treat the classification process as a series of multi-class classification tasks, where each task represents an attempt to classify abnormal traces into specific fault categories for a MSS. TraFaultDia is trained on several abnormal trace classification tasks with a few labeled instances from a MSS using a meta-learning approach. After training, TraFaultDia can quickly adapt to new, unseen abnormal trace classification tasks with a few labeled instances across MSS. TraFaultDia’s use cases 
are scalable depending on how fault categories are built from anomalies within MSS. We evaluated TraFaultDia on two representative MSS, TrainTicket and OnlineBoutique, with open datasets. In these datasets, each fault category is tied to the faulty system component(s) (service/pod) with a root cause.  Our TraFaultDia automatically classifies abnormal traces into these fault categories, thus enabling the automatic identification of faulty system components and root causes without manual analysis. Our results show that, within the MSS it is trained on, TraFaultDia achieves an average accuracy of 93.26\% and 85.20\% across 50 new, unseen abnormal trace classification tasks for TrainTicket and OnlineBoutique respectively, when provided with 10 labeled instances for each fault category per task in each system. In the cross-system context, when TraFaultDia is applied to a MSS different from the one it is trained on, TraFaultDia gets an average accuracy of 92.19\% and 84.77\% for the same set of 50 new, unseen abnormal trace classification tasks of the respective systems, also with 10 labeled instances provided for each fault category per task in each system.

  
  

\end{abstract}

\ccsdesc[500]{Computer systems organization~Maintainability and maintenance}
\ccsdesc[500]{Computer systems organization~Reliability}
\ccsdesc[500]{Software and its engineering~Software maintenance tools}
\keywords{software, microservice, fault category, root cause analysis, trace, meta learning, NLP}


\maketitle
\section{Introduction}
\label{sec:introduction}
Microservice architecture is a software design approach where software systems are developed as a collection of small, independent services that interact via lightweight mechanisms like HTTP APIs~\cite{zhou2018fault}. In a microservice-based system (MSS), a user request can initiate a sequence of interactions among multiple services. Due to such complexity and dynamic nature,  MSS may fail in various fault types, e.g.,  service/server configuration, interaction, message sequence, resource allocation related errors \cite{zhou2018fault}. Traces, which map the path of a user request, are fundamental to understanding and monitoring MSS~\cite{li2022enjoy, OpenTelemetry}. 

Advanced by existing Artificial Intelligence for IT Operations (AIOps) methods, abnormal traces can be automatically detected and responsible services can be automatically identified. Substantial efforts (e.g., \cite{zhang2022deeptralog, chen2023tracegra, zhang2022putracead, chen2022microegrcl,raeiszadeh2023real}) construct trace graphs to capture the complex interactions among services  in MSS. Trace graphs have proven effective not only in detecting abnormal traces but also in providing ranked lists of the potential responsible service(s) for each abnormal trace. The study \cite{yu2023nezha} uses event graphs to map relationships among multimodal monitoring data (including traces) to localize faulty code regions and resource types (e.g., CPU). Besides, some studies \cite{nedelkoski2019anomaly,kohyarnejadfard2022anomaly,du2023trace} treat traces as sequences of service instances/logs and use long short-term memory (LSTM) networks to model these sequences for detecting abnormal traces.

However, even with the use of existing AIOps methods, handling failures of MSS would still require human efforts from practitioners for further root cause analysis (RCA) to diagnose fault categories and analyze failure reasons for detected abnormal traces \cite{zhou2018fault,yu2023nezha,wang2021promises,chen2021trace}. While trace/event graphs that locate potential services/code regions responsible for abnormal traces provide a useful starting point for RCA, they fall short of directly pinpointing the root causes \cite{nguyen2022survey}. Conducting further RCA demands that practitioners possess a deep understanding of software architecture, operational behaviors, and failure modes to effectively analyze and classify various categories of abnormal traces \cite{zhou2018fault,yu2023nezha,nedelkoski2019anomaly2}, particularly when some anomalies may not stem directly from specific service(s)/code regions within MSS. For example, misconfigurations in virtual environments can result in inefficient resource utilization and conflicts between different services in a MSS; failures in third-party libraries may cause dependent MSS's services to fail; high user request loads during peak hours may overwhelm system resources, leading to multiple service failures within MSS \cite{gan2019open,zhou2018fault}. As MSS grow increasingly complex, the volume of trace-related data and fault cases expand, making it infeasible for practitioners to efficiently perform RCA on a large number of detected abnormal traces. Prompt detection of abnormal traces is crucial, but without automation for further RCA, detected abnormal traces can not be addressed in time, may lead to delayed resolutions, increased downtime, and unexpected operational costs \cite{zhou2018fault,ikram2022root}. This highlights a gap in the current AIOps context. 

\textit{In this paper, to narrow the above gap, we propose a novel AIOps framework, TraFaultDia,
to automatically classifying abnormal traces into fault categories for MSS. We treat the classification process as a series of multi-class classification tasks, where each task represents an attempt to classify abnormal traces into specific fault categories for a MSS. }TraFaultDia's use cases are scalable depending on how fault categories are built from anomalies within MSS. For example, in our study, we use open datasets for two representative benchmark MSS, Trainticket and OnlionBoutique. In these open datasets, each fault category is tied to the faulty system component(s) (service/pod) with a root cause, as detailed in Section \ref{sec:emprical_study} and Table \ref{tab:dataset_fault_category}. Our TraFaultDia automatically classifies abnormal traces into these fault categories, thus enabling the automatic identification of faulty system components and root causes without manual analysis. For example, referring to fault categories in our fault dataset (Table \ref{tab:dataset_fault_category}), for OnlineBoutique, an abnormal trace automatically classified as fault category ``B29.adservice'' indicates that this anomaly is associated with the ``adservice pod'' due to the root cause ``Exception code defects''; for Trainticket, an abnormal trace classified as ``F6.SQL error'' indicates that this anomaly is caused by the root cause of the SQL error of a dependent service. This automatic identification of faulty system components and root causes can allow practitioners to quickly understand the nature of failures and their potential impact without the need for extensive manual analysis of each abnormal trace. This simplifies the process of handling a large number of abnormal traces~\cite{chen2021trace}, allowing practitioners to prioritize them based on categorized fault type and severity. It can lead to more targeted and efficient resource allocation: the right teams or tools can be promptly deployed to tackle specific abnormal traces, ensuring that expertise and resources are utilized optimally and not wasted on unsuitable tasks \cite{zhou2018fault,nedelkoski2019anomaly2,yu2023nezha}. This not only speeds up the resolution process but also enhances the overall operational efficiency. However, through our empirical study on Trainticket and OnlionBoutique, and their trace-related data in open datasets in Section \ref{sec:emprical_study}, we identified three significant challenges associated with abnormal trace classification tasks across MSS: C1.MSS heterogeneity, C2.high dimensional, multi-modal trace-related data, C3.imbalanced abnormal trace distribution in fault categories. Further details are provided in Section \ref{sec:emprical_study}. 

We design TraFaultDia to meet our aim and address challenges C1-C3. TraFaultDia is trained on several abnormal trace classification tasks with a few labeled instances using a meta-learning approach. This enables it to quickly adapt to new, unseen abnormal trace classification tasks with a few labeled instances (C3) for any MSS (C1) after training. To represent abnormal traces in a MSS, TraFaultDia constructs trace representations by fusing high-dimensional, multi-modal trace-related data (C2) and compressing them into low-dimensional embeddings, ensuring both effective and efficient abnormal trace classification. We evaluate our TraFaultDia through several experiments
on TrainTicket and OnlineBoutique, with open datasets. We define two research questions:
\begin{itemize}
    \item  \textbf{RQ1: Within-system adaptability}. How effectively and efficiently can TraFaultDia, once trained on abnormal trace classification tasks within a MSS, adapt to new abnormal trace classification tasks within the same MSS?
    
    \item  \textbf{RQ2: Cross-system adaptability.} How effectively and efficiently can TraFaultDia, once trained on abnormal trace classification tasks within a MSS, adapt to new abnormal trace classification tasks in a different MSS?
\end{itemize}


\textbf{Significance.} RQ1 evaluation is critical because within-system adaptability allows the framework to maintain its effectiveness in categorizing abnormal traces in the changing context within a MSS. Since MSS are dynamic with frequent service updates, additions, or removals, new abnormal traces from novel fault categories may appear \cite{zhou2018fault}. A framework that can adapt to new abnormal trace classification tasks within a MSS without extensive retraining would significantly reduce the costs and rework effort while increasing practical utility in practices. RQ2 evaluates our framework's capability to transfer learned knowledge from one MSS to other MSS. Prior studies \cite{chen2020logtransfer,zhang2024metalog,han2021unsupervised, wang2024cross} have investigated cross-system adaptability of AIOps methods for anomaly detection across software systems. In our study context, cross-system adaptability is valuable for organizations that run several MSS, as it allows them to use the same framework for abnormal trace classification tasks across different MSS without extensive training on each MSS; it also can allow organizations to train the framework using academic benchmark MSS open datasets and rapidly adapt this framework to industry MSS with only a few labeled trace instances for effective abnormal trace categorization. Our main \textbf{contributions} are highlighted as follows:


\begin{itemize}
    \item We present TraFaultDia, an AIOps framework that automatically classifies abnormal traces into specific fault categories across various MSS. It requires only a few labeled abnormal trace instances from the target MSS 
    making it efficient and practical for real-world applications. 
    
    \item We employ an unsupervised approach to fuse high-dimensional, multi-modal trace-related data into compressed yet effective trace representations, facilitating 
    efficient and effective subsequent trace analysis.
    
    \item Evaluation results on representative benchmark MSS with open datasets demonstrate the performance of our approach.

\end{itemize}
\section{background}
\subsection{Trace structure}
\begin{figure}[ht]
    \centering
    \begin{minipage}{0.45\textwidth}
        \centering
        \includegraphics[width=0.5\textwidth]{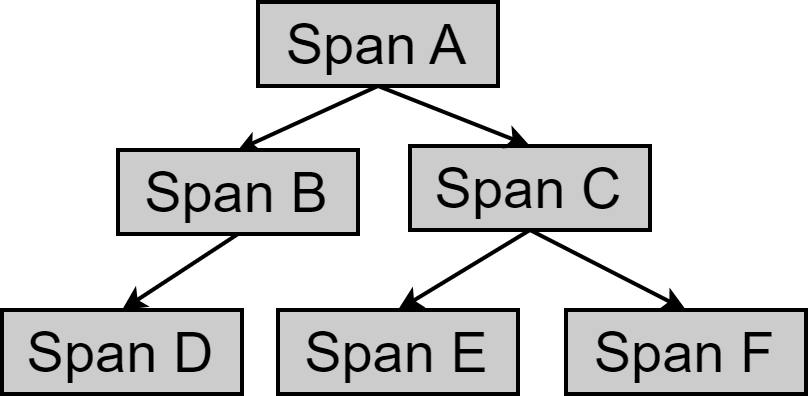} 
        \caption{An example trace structure (Zhang et. al.~\cite{zhang2022deeptralog})}
        \label{fig:TraceTree}
        \Description{A trace tree showing the structure of an example trace}
    \end{minipage}\hfill
    \begin{minipage}{0.5\textwidth}
        \includegraphics[width=0.9\textwidth]{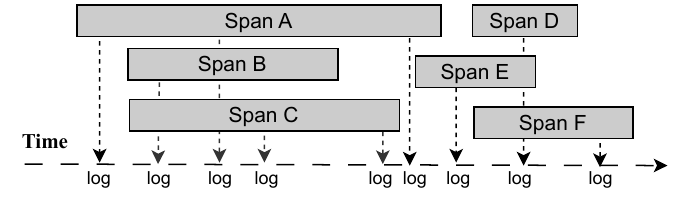} 
        \caption{Spans and logs in the timeline (modified from Zhang et. al.~\cite{zhang2022deeptralog})}
        \label{fig:TraceTimeline}
        \Description{A trace timeline}
    \end{minipage}
\vspace{-10pt}
\end{figure}

Based on OpenTelemetry \cite{OpenTelemetry}, a trace is structured as a hierarchical tree of spans. Each span is an individual operation performed by a particular service. The root span is the starting point of the trace, akin to the tree's base, from which all other spans branch out. Each span has a parent span, except for the root span. The span that is currently being executed (i.e., active span) may contain nested sub-spans, which represent smaller units of work that are part of the larger operation encompassed by the active span to which they belong. 
Figure~\ref{fig:TraceTree} shows an example tree structure for a trace, where Span A is the root span, and it triggers a sequence of calls to other spans. Figure~\ref{fig:TraceTimeline} shows the same spans depicting how a request flows through the execution of each span in sequence and pinpoints the time point where relevant logs are generated. Logs record the behaviors of service instances in spans.

\subsection{Empirical study on MSS and their traces in open datasets} \label{sec:emprical_study}
To evaluate our framework, we use two open datasets: DeepTraLog \cite{DeepTraLog} and Nezha \cite{Nezha}. We empirically observed two benchmark MSS, TrainTicket and OnlineBoutique, and their traces in DeepTraLog and Nezha. DeepTraLog includes 
normal traces, and abnormal traces across 14 fault categories in Trainticket. These fault categories span asynchronous interaction, multi-instance, configuration, and monolithic dimensions, and are designed to replicate real-world anomaly scenarios. Nezha comprises normal traces and abnormal traces from both TrainTicket and OnlineBoutique. It includes abnormal traces caused by five fault types—CPU contention, CPU consumption, network delay, error return, and exception code defect—each fault type was applied to various service pods within each system. In this context, each pod associated with a fault type represents a unique fault category.



For our abnormal trace study, we established our fault dataset using DeepTraLog and Nezha. Our fault dataset includes abnormal traces from 30 fault categories (F1-F30) for TrainTicket and 32 fault categories (B1-B32) for OnlineBoutique. Table~\ref{tab:dataset_fault_category} summarizes fault categories in each system, while Table~\ref{tab:fault_dataset_statistics} provides descriptive statistics on abnormal traces in these fault categories within each system. Through observations, we found three challenges for abnormal trace categorization across MSS. We use TrainTicket and OnlineBoutique, along with our fault dataset, as examples to describe these three challenges below.

\begin{table} [ht]
\centering
\caption{Fault categories of TrainTicket and OnlineBoutique in our fault dataset.}
\small
\label{tab:dataset_fault_category}
\begin{tabular}{ p{13.5cm}}
\toprule
   \textbf{TrainTicket} \\
 DeepTraLog: \textit{Asynchronous service invocations related faults} (F1.Asynchronous message sequence error, F2.Unexpected order of data requests, F13.Unexpected order of price optimization steps);\textit{ Multiple service instances related faults} (F8.Key passing issues in requests, F11.BOM data is updated in an unexpected sequence, F12.Price status query ignores expected service outputs); \textit{Configuration faults} (F3.JVM and Docker configuration mismatch, F4.SSL offloading issue, F5. High request load, F7. Overload of requests to a third-party service); \textit{Monolithic faults} (F6.SQL error of a dependent service, F9.Bi-directional CSS display error, F10.API errors in BOM update, F14.Locked product incorrectly included in CPI calculation) \\
 
 Nezha: \textit{CPU contention} on F23.travel, F25.contact, F26.food service pods; \textit{Network delay} on F28.basic, F29.travel, F30.route, F27.security, F24.verification-code service pods; \textit{Message return errors} on F16.basic, F15.contact, F18.food, F19.verification-code service pods; \textit{Exception code defects} on F17.basic, F21.route, F22.price, F20.travel service pods.
\\ \hline

 \textbf{OnlineBoutique}  \\
 Nezha: \textit{CPU contention} on B4.shipping, B14.cart, B18.currency, B19.email, B26.recommendation, B31.adservice, B9.payment, B11.frontend service pods; \textit{CPU consumption} on B8.recommendation, B12.frontend, B24.productcatalog, B28.shipping, B17.checkout, B20.email, B32.adservice service pods; \textit{Network delay} on B10.currency, B1.cart, B15.checkout, B22.productcatalog, B27.shipping, B21.payment, B25.recommendation, B29.adservice, B7.email service pods; \textit{Message return errors} on B6.frontend, B23.productcatalog, B2.checkout, B30.adservice  service pods; \textit{Exception code defects} on B5.adservic, B13.frontend,  B3.productcatalog,  B16.checkout   service pods  \\
\bottomrule
\end{tabular}
\vspace{-5pt}
\end{table}

\begin{table}[ht]
\small
\centering
\caption{Descriptive statistics on traces in our fault dataset.}
\label{tab:fault_dataset_statistics} 
\begin{minipage}[b]{0.48\linewidth}\centering
\begin{tabular}{p{3cm}rrr}
\toprule
\textbf{TrainTicket} & Mean & Min & Max  \\
\midrule
Unique traces per &  & &  \\
fault category: & 1196  & 26 & 2546 \\ \hline
Spans per trace: & 79 & 1 & 345 \\ \hline
Logs per trace: & 44 & 1 & 340 \\
\hline
\end{tabular}
\end{minipage}
\hfill
\begin{minipage}[b]{0.48\linewidth}\centering
\begin{tabular}{p{3cm}rrr}
\toprule
\textbf{OnlineBoutique} & Mean & Min & Max  \\
\midrule
Unique traces per &   &  &   \\
fault category: &  443 & 32 & 1018  \\ \hline
Spans per trace: & 53 & 1 & 190 \\ \hline
Logs per trace: & 51 & 4 & 184 \\ 
\hline
\end{tabular}
\end{minipage}
\end{table}

 \textit{C1.MSS heterogeneity.} MSS vary significantly, each composed of distinct services with unique behaviors. This heterogeneity challenges the development of a universal approach for trace representation and classification across various MSS. For instance, TrainTicket, a train ticket booking MSS, consists of 45 services, while OnlineBoutique, an e-commerce platform, has 12 services. Their distinct system nature and service compositions contribute to different system behaviors.

  \textit{C2. High dimensional, multi-modal trace-related data.} 
  This complexity makes it impractical to rely solely on raw trace-related data for constructing representations. Fusing complex trace-related data into the compressed representation is essential for effective and efficient abnormal trace classification in MSS. Referring to Table \ref{tab:fault_dataset_statistics}, we observe trace-related data complexity: in each trace, the number of spans and logs varies greatly, from one to hundreds. Upon further examination of spans and logs from these traces, and referring to industry standards from OpenTelemetry \cite{OpenTelemetry}, we found that spans and logs reveal their multi-modal nature: spans have textual (call component and path), time-based (span start and end time), and identity (trace ID, span ID) attributes; logs have textual (log component and message, severity level) and identity (trace ID, span ID) attributes. Each modality captures distinct, critical operational information for MSS. Missing any modality would impact the performance of our framework to classify abnormal traces into fault categories:
  

  \begin{itemize}
    \item Trace IDs in spans and logs are used to identify which traces they belong to. 
    \item Span textual attributes, ``call component and path'', are crucial for recognizing abnormal traces from most fault categories in our fault dataset (Table~\ref{tab:dataset_fault_category}).
    These attributes provide critical insights into service operations: for each span, the ``call component'' indicates the part of a system involved in a service call, and the ``call path'' is the route a service call takes. 
    
    \item Span time-based attributes, span start and end time, reflect the running time of service operations in spans. These attributes are essential for recognizing latency-related abnormal traces caused by, e.g., network delay and CPU contention issues in our fault dataset (Table~\ref{tab:dataset_fault_category}).

    \item Span IDs are designed with a hierarchical structure reflecting the relationship between the active spans and their nested sub-spans \cite{OpenTelemetry}. Take Figure \ref{fig:TraceTimeline} as an instance, assuming that Span~A is the active span, both Span B and Span~C are nested in Span A. When Span~A initiates, it is assigned a Span ID ``a480f2.0'', while its nested spans, Span B~and Span C,  are assigned with the derivative Span IDs ``a480f2.1'' and ``a480f2.2'' respectively. Spans like Span D and Span E, which are not nested sub-spans of any active spans, receive distinct Span IDs, like ``a343mc.0'' and ``a987gq.0'', to reflect their separate execution pathways. The hierarchical structure of Span IDs would be useful in identifying faults that involve the interaction and sequencing of multiple services, e.g., asynchronous service invocations, and multiple service instances related faults in our fault dataset (Table~\ref{tab:dataset_fault_category}).
    
    \item For log textual attributes, ``log component'' states the part of the system that generates the log message, ``log messages'' are written by developers reflecting the operation state,  ``severity level'' (e.g., INFO, WARN, ERROR) indicates the urgency or importance of log messages. These attributes are crucial for recognizing abnormal traces in fault categories where detailed status information about service operations is essential for diagnosing issues, e.g., code, message return, and configuration errors in our fault dataset~(Table~\ref{tab:dataset_fault_category}).
  \end{itemize}

 \textit{C3. Imbalanced abnormal trace distribution in fault categories.} In MSS, certain fault categories may have very few sampled abnormal traces compared to others, e.g., as shown in Table \ref{tab:fault_dataset_statistics}, some fault categories only have around 30 instances while others have over thousands for both TrainTicket and OnlineBoutique. This imbalance may arise when some fault categories are rarer than others. It poses a challenge for training and evaluating classification models on fault categories where abnormal trace samples are limited.

\subsection{Related work}
\label{sec:sec_related_work}

\paragraph{Trace representation} In MSS trace analysis, many studies (e.g., \cite{zhang2022deeptralog, chen2023tracegra, zhang2022putracead, chen2022microegrcl,raeiszadeh2023real}) utilized GNNs to build trace graphs. Trace graphs are constructed by modeling spans as nodes and their interactions as edges to reflect the flow of requests within traces. Nodes are detailed using span attributes (e.g., call path, call response time), and/or by incorporating textual attributes from logs associated with the spans. Trace graphs have demonstrated effectiveness in detecting abnormal traces and locating potential fault-causing services. However, they are not suitable for our case due to their computational expense and scalability issues \cite{waikhom2023survey}. For example, in our fault dataset, some traces consist of hundreds of spans, each span potentially connected to many others within these traces, see Table \ref{tab:fault_dataset_statistics}. This setup exponentially increases the complexity of modeling dependencies among spans to edges for building trace graphs, leading to high computational costs. In real-world MSS, traces often contain even more spans with more complex interactions~\cite{zhang2022deeptralog,dragoni2017microservices}, further increasing computational costs. Moreover, in dynamic environments of MSS, where traces continuously evolve, updating trace graphs to reflect new or altered spans and dependencies is computationally burdensome. This does not fit our needs for efficient adaptability both within and across dynamic MSS. Apart from trace graphs, several anomaly detection studies ~\cite{nedelkoski2019anomaly,kohyarnejadfard2022anomaly,du2023trace} treated a trace as a sequence of spans and used span attributes to construct trace representations. However, these trace representations may be ineffective for recognizing certain fault categories in our fault datasets, as they completely overlook logs as part of traces, as discussed in Section \ref{sec:emprical_study}.

\textit{Trace classification:} Many MSS anomaly detection studies (e.g.,  
\cite{kohyarnejadfard2019system,zhang2022putracead,kong2024grand}) use binary classification to determine whether traces are abnormal or not using two classes of labels: anomaly or normal. Binary classification differs from our work, which involves multi-class classification on detected abnormal traces. In our case, each trace is associated with one of many fault categories, making it a multi-class classification problem. This requires a more nuanced approach to classify each trace into a specific fault category from among many fault categories. Existing studies on this topic are very rare for MSS or similar cloud-based systems. The study~\cite{nedelkoski2019anomaly2} is the only related work. It uses a convolutional neural network (CNN) to classify abnormal traces into four time series-based fault categories: incremental, mean shift, gradual increase, cylinder. It characterizes the trace as a sequence of spans and uses the time-series data on the span attribute ``call path'' to do multi-class classification. This approach is insufficient to address our fault dataset that includes a broader range of fault categories (Table \ref{tab:dataset_fault_category}) and associates with our identified challenges C1\&C2. Specifically, MSS heterogeneity (C1) makes it challenging to train a single CNN model to perform effectively across different MSS. Its approach only considers time-series data on the span attribute ``call path'', while it overlooks other trace-related data modalities (C2) that would affect recognizing abnormal traces from certain fault categories in our fault dataset, as discussed in Section \ref{sec:emprical_study}. 

Our study builds upon existing research and provides novel insights. 
First, to build effective trace representations, our approach incorporates all essential span and log attributes identified through our empirical research on MSS and their traces in open datasets (Section \ref{sec:emprical_study}). We fuse these attributes into the unified,  compressed trace representations, thereby ensuring the effevtinvess and efficiency of trace analysis. Compared to GNN-based methods, our approach does not necessitate frequent updates to reflect dependencies among new or altered spans. This makes our approach more scalable and computationally efficient, providing significant advantages in dynamic and complex environments of MSS. Second, our framework goes beyond prior approaches that focus on binary classification for normal and abnormal traces. We perform multi-class classification to identify specific fault categories for abnormal traces across various MSS environments.

\section{Methodology}
\subsection{Our framework TraFaultDia workflow and design rationale}
\label{sec:ourFrameworkDesign}
\begin{figure}[ht]
\vspace{-0.5cm}
\centering
\includegraphics[width=0.95\textwidth]{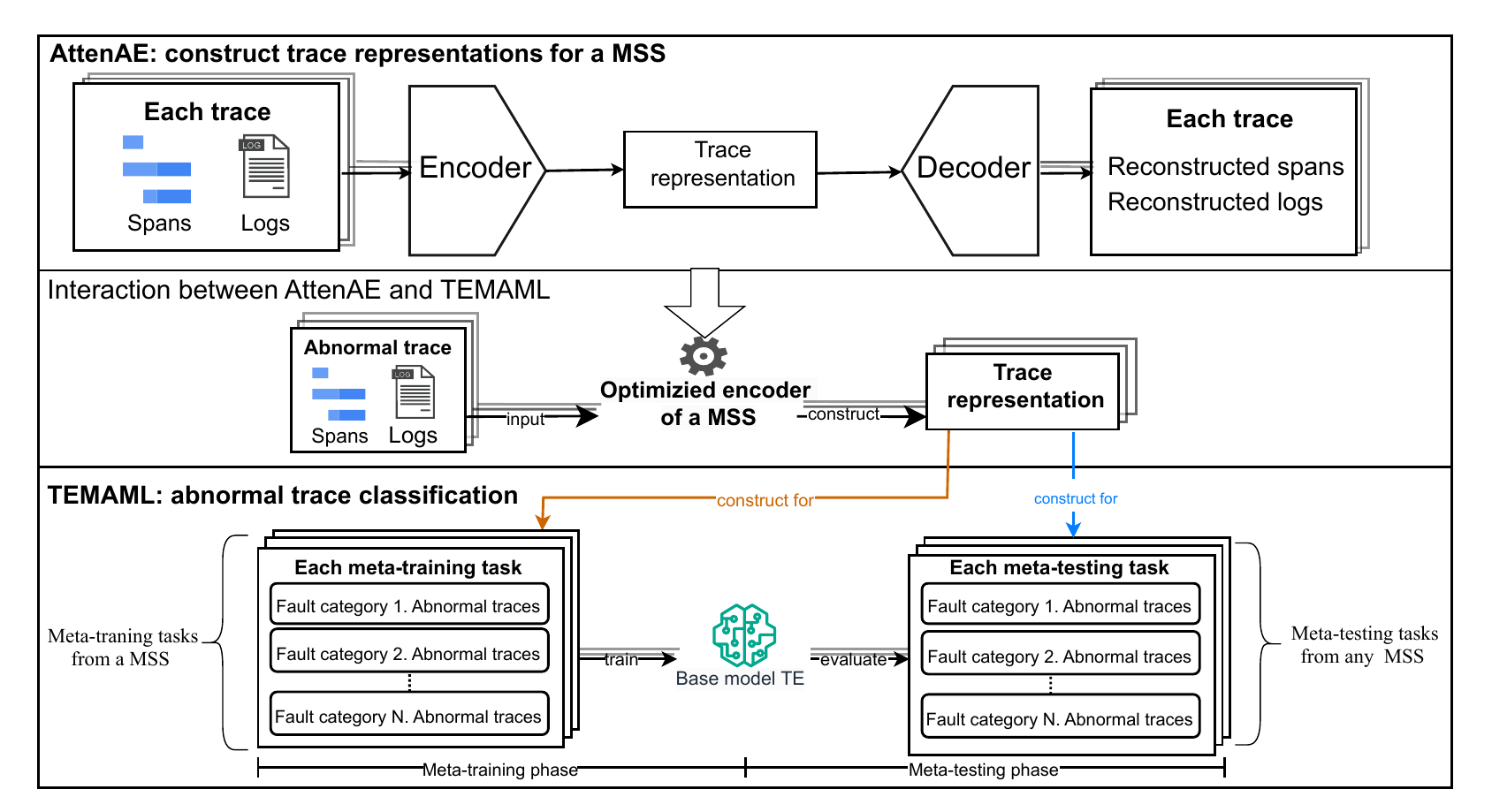}
\vspace{-0.5cm}
\caption{Overview of our framework}
\Description{An overview diagram of our framework, illustrating the key components and their interactions.}
\label{fig:framework_overview}
\vspace{-0.3cm}
\end{figure}

Following the
principle of meta-learning \cite{finn2017model}, we define each abnormal trace classification task from any MSS in the N-Way K-Shot setup. This means that each task involves N distinct fault categories, with each category having K labeled example abnormal traces. Figure~\ref{fig:framework_overview} shows the overview of our framework TraFaultDia. TraFaultDia has two components: the Multi-Head Attention Autoencoder (AttenAE), and the Transformer-Encoder based Model-Agnostic Meta-Learning (TEMAML) model.

\textbf{Our framework workflow:} (1)
Given an MSS, AttenAE is trained on sufficient unlabeled traces to unsupervisedly learn to fuse trace-related data into the compressed yet effective trace representations. AttenAE consists of an encoder and a decoder: the encoder generates trace representations by fusing original trace-related data; the decoder reconstructs the original data from these fused representations. AttenAE is trained by reducing the loss between original trace-related data and reconstructed trace-related data. Once trained, the encoder is utilized independently to generate trace representations for new, unseen traces within this MSS. (2) TEMAML trains the base model, transformer-encoder (TE), to do abnormal trace classification tasks in any MSS. In the meta-training phase, TEMAML is trained on abnormal trace classification tasks from a MSS, referred to as meta-training tasks. In the meta-testing phase, TEMAML is evaluated on new, unseen abnormal trace classification tasks from any MSS, referred to as meta-testing tasks. A meta-training/meta-testing task here refers
to a training episode in meta-learning \cite{finn2017model}, denoting a single iteration where the base model is trained/evaluated on
the MSS in our study. Each meta-training and meta-testing task is in the N-way K-shot setup. Abnormal traces in meta-training/meta-testing tasks come from their respective MSS. TEMAML uses the optimized AttenAE's encoder of a MSS to construct trace representations for representing abnormal traces in this MSS's meta-training and meta-testing tasks. 


\textbf{Design rationale:} 
Our framework is designed to tackle our research questions (Section \ref{sec:introduction}) and identified challenges C1-C3 (Section \ref{sec:emprical_study}). (1) \textit{Why AttenAE}: We use AttenAE because it serves as an autoencoder, capable of fusing raw high-dimensional, multimodal trace-related data (C2) into unified, low-dimensional representations, which are essential for effective and efficient trace analysis. 
The multi-head attention mechanism within AttenAE recognizes and integrates the most relevant features of trace-related data for constructing trace representations. Also, AttenAE supports unsupervised training, allowing our framework to learn from unlabeled traces that are easily obtainable from MSS.  AttenAE has been widely used in AI studies (e.g., \cite{zhou2020tafa,chen2020transformer,huang2020multimodal}) to fuse high-dimensional and multi-modal data, such as images, text, and audio, into unified, compressed representations unsupervisedly. (2) \textit{Why TEMAML}: 
We use TE as the base model to classify abnormal traces (represented by trace representations constructed by AttenAE) into precise fault categories. Since these trace representations are latent representations that are a fusion of trace-related data and not true sequences, sequence models may not be appropriate. 
TE excels at recognizing and integrating the most important features from such latent trace representations, leveraging its self-attention mechanism \cite{vaswani2017attention}, making it the most suitable choice for our case.  We chose MAML algorithm \cite{finn2017model} to train TE as it can provide TE with few-shot learning capability to address C3 (by recognizing abnormal traces from both frequent and rare fault categories with just a few labeled instances) and transfer learning capability to address C1 (by enabling within- and across-system adaptability, even with heterogeneous MSS). MAML has been used for few-shot learning and transfer learning in practical contexts. For instance, the studies~\cite{zhang2024metalog,wang2024cross} used it to train a sequence model on source systems with sufficient labeled log data and then adapted it on target systems with fewer labeled logs for software log anomaly detection; MAML has been applied to medical imaging tasks to enable models to learn from small, labeled datasets and quickly adapt to new and rare disease category tasks with a few labeled examples \cite{maicas2018training}.

\subsection{AttenAE for constructing trace representations}
\label{sec_trace_representation_gen}

\begin{figure}[ht]
\centering
\includegraphics[width=1\textwidth]{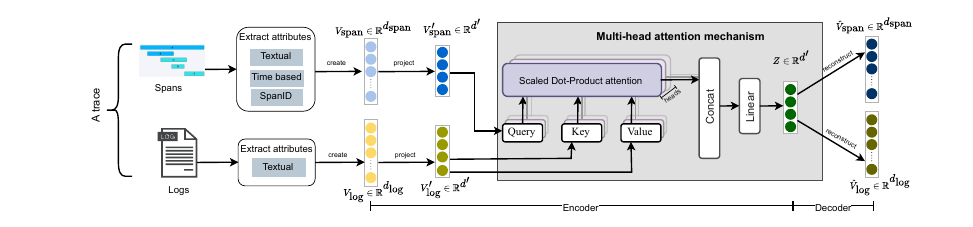}
\vspace{-0.8cm}
\caption{AttenAE architecture}\Description{fusion framework}
\label{fig_AEFusion}
\vspace{-8pt}
\end{figure}

Figure \ref{fig_AEFusion} shows the structure of AttenAE. For a given MSS, we denote a set of traces as $Tr = \{Tr_1, Tr_2,..., Tr_n\}$, where each $Tr_{\text{i}}$
represents an individual trace consisting of a sequence of spans and logs. $Tr =(Span, Log)$ represents the combination of spans and logs across all traces.

\subsubsection{Span prepossessing and vector generation}
\label{sec:spanembedding} 
For each span, we extract all textual (call component and path), time-based (span start and end time), and identity (trace ID, span ID) attributes explored in Section \ref{sec:emprical_study}. %
We use trace IDs to collect spans into their belonging traces. We normalize time-based attributes (in UNIX format) within each span's context, considering unique characteristics and scale of each span. We concatenate time-based attributes for each span, obtaining a single vector $V_{\text{numeric}}$ for $Span$ of $Tr$. For span IDs, we abstract away the shared common prefix in span IDs and only retain hierarchical-level digits. Taking our example explaining the hierarchical structure of spans IDs in Section \ref{sec:emprical_study}, we reassign span IDs for Span A, Span B, Span C, Span D, and Span E from ``480f2.0, 480f2.1, a480f2.2, a343mc.0, a987gq.0'' to ``1.0, 1.1, 1.2, 2.0, 3.0''. We normalize span IDs within the trace context, resulting a vector $V_{\text{span\_id}}$ for $Span$ of $Tr$.

We concatenate textual attributes (call component and path) to form a singular attribute termed ``service operation''. Template-based representation \cite{he2017drain} of textual attributes may not be ideal for our observed MSS, where diverse service operations could result in numerous templates and they keep changing and include many out-of-vocabulary (OOV) words \cite{le2021log}. Thus, we employ a neural representation method \cite{le2021log} that captures the semantics of service operations using the pre-trained model BERT with the subword technique to better handle new, changing, OOV words in evolving service operations. To construct a neural representation for service operations, we undertake three steps. \textit{Step 1. Prepossessing.} We convert all uppercase letters to lowercase, substitute specific variables with standard identifiers (e.g., replace instances like ``Prod1234'' with ``ProductID''), and remove any non-alphabetic characters. \textit{Step 2. Tokenization.} We use WordPiece tokenization~\cite{wu2016google} to tokenize service operations into subwords. 
\textit{Step 3. Neural representation.} We feed the subwords, into the BERT base model \cite{bert2018} to generate word embeddings for each sub-word. We use word embeddings generated by the last encoding layer of the model and calculate the sentence embedding of each service operation as the average of its word embeddings. This process yields a vector representation $V_{\text{operation}}$ for service operations of $Span$ of $Tr$.

For $Tr$, we concatenate vector representations acquired from the preceding phase, thereby establishing a composite vector $V_{\text{span}} \in \mathbb{R}^{d_{\text{span}}}$
for $Span$, where $d_{\text{span}}$ represents the dimensionality of the vector space for $V_{\text{span}} = \text{Concat}(v_{\text{numeric}}, v_{\text{span\_id}}, v_{\text{operation}} )$


\subsubsection{Log prepossessing and vector generation} We extract textual (log component and message, severity level) and identify (trace ID) attributes from logs. We use trace IDs to collect logs into their belonging traces.  We concatenate these attributes to form a singular attribute ``log event''. We build neural representations for log events omitting the step of log parsing \cite{mantyla2024loglead}. 
The empirical studies \cite{le2021log, hashemi2021onelog} observed that log parsing might provide limited benefits in dynamic software systems with evolving logs.  We also use the neural representation method, as we did for service operations, to capture log events' semantics, since log events are also diverse, contain many OOV words, and are continuously evolving in MSS.
Our method for building neural representations of log events follows the same three-step process as the one that is used for service operations in spans in Section \ref{sec:spanembedding}. This process yields a vector representation $V_{\text{log}} \in \mathbb{R}^{d_{\text{log}}}$, which comprises sentence embeddings of log events, for $Log$ of $Tr$. Here, $d_{\text{log}}$ represents the dimensionality of the vector space for $V_{\text{log}}$.






\subsubsection{Trace representation construction} 

For a given MSS, we construct trace representations for traces $Tr$ utilizing our AttenAE's encoder. The encoder first projects the input vectors $V_{\text{span}}$ and $V_{\text{log}}$ into a common feature space~$\mathbb{R}^{d'}$:

\begin{equation}
V'_{\text{span}} = g(W_{\text{span}}V_{\text{span}} + b_{\text{span}}) \text{;}  \quad 
V'_{\text{log}} = g(W_{\text{log}}V_{\text{log}} + b_{\text{log}})
\end{equation}
here, \( g \) denotes the activation function, \( W_{\text{span}} \) and \( W_{\text{log}} \) are respective weight matrices, and \( b_{\text{span}} \) and \( b_{\text{log}} \) are the bias vectors. AttenAE's encoder incorporates the multi-head attention mechanism~\cite{vaswani2017attention}:



\begin{equation} 
\label{eq:multi-headAttn}
\left\{
    \begin{array}{lr}
    \text{Attention}(Q, K, V) = \text{softmax}\left(\frac{QK^T}{\sqrt{d_k}}\right)V,  \\
    \text{head}_i = \text{Attention}(QW^Q, KW^K, VW^V),\\
    \text{MultiHead}(Q, K, V) = \text{Concatenate}(\text{head}_1, \ldots, \text{head}_h)W^O &  
    \end{array}
\right.
\end{equation}
where Q, K, and V refer to the query, key, and value matrices, respectively. This mechanism computes initial attention scores by taking the dot product of Q and K, regulates these scores by ${\sqrt{d_k}}$ for numerical stability, and then applies a softmax function to produce the attention distribution.  This distribution assigns weights to the elements in V. Each attention head ($\text{head}_i$) is computed through separate learned projections of Q, K, and V using the matrices ${W}^Q$, ${W}^K$, ${W}^V$ as learnable weights. It results in distinct Q, K, and V for $\text{head}_i$. The final multi-head attention output is created by concatenating the outputs of all individual attention heads into a single vector and linearly transforming it using weight matrix ${W}^O$. Our AttenAE's encoder takes \( V'_{\text{span}} \) and \( V'_{\text{log}} \) as input into the above mechanism to fuse them into  the trace representations Z for traces $T_r$:
\begin{equation}
    \text Z = {MultiHead}(V'_{\text{span}}, V'_{\text{log}}, V'_{\text{log}})
\end{equation}

\noindent where, we set \( V'_{\text{span}} \) as Q, and \( V'_{\text{log}} \) as both K and V. 
This setup aligns with the roles of spans in reflecting the trace structure and service communications, while logs provide detailed contextual event information \cite{OpenTelemetry}. Thus, 
for a set of traces $Tr = \{Tr_1, Tr_2,..., Tr_n\}$, we generate the corresponding trace representations $Z = \{Z_1, Z_2,..., Z_n\}$, where $Z_i$ corresponds to $Tr_{\text{i}}$. Our AttenAE's decoder reconstructs the trace representations Z into the original span and log vectors, effectively inverting the encoder's process: \begin{equation}
\begin{array}{lr}
    \hat{V}_{\text{span}} = g(W'_{\text{span}}Z + b'_{\text{span}}); \quad 
    \hat{V}_{\text{log}} = g(W'_{\text{log}}Z + b'_{\text{log}})
\end{array}
\end{equation}


\noindent where $\hat{V}_{\text{span}} \in \mathbb{R}^{d_{\text{span}}}$, $\hat{V}_{\text{log}} \in \mathbb{R}^{d_{\text{log}}}$, \( g \) is the activation function, \( W'_{\text{span}} \) and \( W'_{\text{log}} \) are the respective weight matrices, and \( b'_{\text{span}} \) and \( b'_{\text{log}} \) are the bias vectors. Training AttenAE includes optimizing its parameters $\Psi$ to minimize the overall loss $\mathcal{L}$ between original $(V_{\text{span}}, V_{\text{log}})$ and their respective reconstructed vectors $(\hat{V}_{\text{span}}, \hat{V}_{\text{log}})$: 
$\begin{array}{lr}
    \min_{\Psi} \mathcal{L} = \lVert \hat{V}_{\text{span}} - V_{\text{span}} \rVert^2 + \lVert \hat{V}_{\text{log}} - V_{\text{log}} \rVert^2 
\end{array}$


\subsection{TEMAML for few-shot abnormal trace classification across MSS}
\label{sec_TE-MAML_method}
\begin{figure}[ht]
\centering
\includegraphics[width=0.8\textwidth]{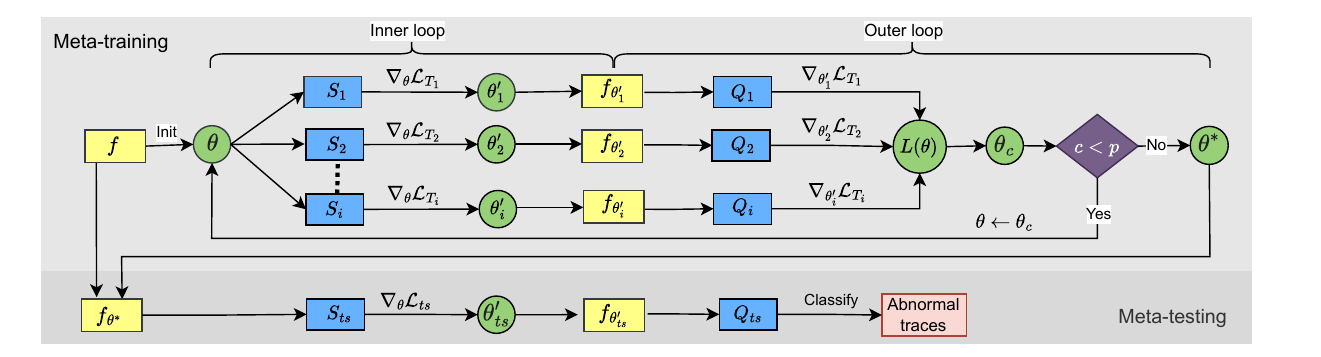}
\caption{TEMAML learning process}
\label{fig:maml}
\vspace{-8pt}
\Description{TEMAML learning process}
\end{figure}

Figure \ref{fig:maml} illustrates TEMAML's basic architecture. TEMAML progresses through two phases: meta-training (where TE is trained using meta-training tasks), and meta-testing (where TE is adapted and evaluated on meta-testing tasks). TEMAML's base model TE, and each phase are explained~below.

\subsubsection{Base model for abnormal trace classification} 
TEMAML trains the base model TE, denoted as~$f$,  to perform multi-class classification for abnormal traces. For an abnormal trace classification task from a MSS, TE operates through the following workflow. First, TE receives trace representations $Z$ as input, where $Z$ represents all abnormal traces $Tr$ in this task. Second, the input $Z$ is processed through TE's self-attention mechanism, which weighs the most relevant parts and captures dependencies within each trace representation $Z_i$ for recognizing the fault type for each trace $Tr_{\text{i}}$. 
This mechanism adheres to the multi-head attention mechanism in Equation \ref{eq:multi-headAttn}. It is called ``self-attention'' since it uses the same input $Z$ as $Q, K, V$ in the attention process: $output = {MultiHead}(Z, Z, Z)$. Third, the output from the self-attention mechanism is further passed through a pooling layer to highlight key features, a dropout layer to prevent overfitting, and a fully connected layer that reshapes the refined output into a suitable format for classification. Finally, a softmax classifier processes the output from the fully connected layer to compute probabilities for each fault category.

\subsubsection{Meta-training} This phase aims to train TE to find robust parameters that can quickly adapt to abnormal trace classification tasks from any MSS. We train TE on several abnormal trace classification tasks (meta-training tasks), denoted as $T = (S, Q)$, which are sampled from a MSS. Each meta-training task is unique and denoted by $T_i = (S_i, Q_i)$, where $S_i$ is a support set and $Q_i$ is a query set for the $i$-th task. The support set $S_i = \{ (z_{ij}^{spt}, y_{ij}^{spt}) \}_{j=1}^{N \times K}$ is indexed by j from 1 to $N \times K$. Here, $N \times K$ follows our N-way K-shot setup, indicating that there are $N$ distinct fault categories and each category has $K$ labeled trace instances. Each $(z_{ij}^{spt}, y_{ij}^{spt})$ is a pair of a trace representation and its corresponding fault category label. 
Similarly, the query set $Q_i = \{ (z_{ig}^{qry}, y_{ig}^{qry}) \}_{g=1}^{N \times M}$, indexed
by g from 1 to $N \times M$. Here, $N \times M$ indicates there are N distinct fault categories and
each has M labeled trace instances. $M$ is greater than $K$ (i.e., $|Q_i| > |S_i|$) to ensure robust optimization across all meta-training tasks. Trace representations $z_{ij}^{spt}$ and $z_{ig}^{qry}$ for $S_{i}$ and $Q_{i}$ respectively are generated using the optimized AttenAE's encoder for a given MSS.

As shown in Figure \ref{fig:maml}, in the meta-training phase, we train our base model TE ($f$) in two loops: an inner loop and an outer loop. The inner loop is responsible for task-level learning, wherein $f$ is adapted to each meta-training task $T_i$. The outer loop optimizes $f$'s parameters to ensure that a few gradient steps yield optimal performance across all tasks $T$. \textit{In the inner loop}, we first randomly initialize $f$'s parameters, represented as a parameterized $f_\theta$ with parameters $\theta$. When adapting to each meta-training task~$T_i$,$f$’s parameters $\theta$ are transformed into task-specific parameters $\theta'_i$, corresponding to the updated model $f_{\theta'_i}$. $\theta'_i$ is computed using gradient descent updates on the support set $S_i$ of $T_i$. Each gradient descent update is computed as:

\begin{equation}
\label{eq:innerUpdate}
\theta'_i = \theta - \alpha \nabla_{\theta} \mathcal{L}_{T_i}(f_{\theta}(S_i))
\end{equation}

\noindent where $\alpha$ is the learning rate for inner loop updates and $\mathcal{L}_{{T}_i}$ is the loss on $S_i$ of ${T}_i$. \textit{In the outer loop}, $f$'s parameters are trained by optimizing the performance of $\theta'_i$ with respect to $\theta$ across all meta-training tasks $T$. It minimizes the overall loss $\mathcal{L}$ on query sets $Q$ of all meta-training tasks $T$:  

\begin{equation}
\min_{\theta} \mathcal{L}(\theta) = \sum_{T_i \in T} \mathcal{L}_{T_i}(f_{\theta'_i}(Q_i))
\end{equation}
 where $\mathcal{L}_{T_i}$ is the loss on $Q_i$ of $T_i$. $f$' parameters $\theta$ are updated p times, where p is the predetermined number of optimization steps. Each update gets the updated parameters $\theta_c$. For example, when using one gradient update (i.e., p = 1), 

\begin{equation}
\theta \leftarrow \theta - \beta \nabla_{\theta} \sum_{T_i \in T} \mathcal{L}_{T_i}(f_{\theta'_i}(Q_i))
\end{equation}

\noindent where $\beta$ is the learning rate for outer loop updates. The standard MAML outer loop updates involve a computationally intensive process of computing a gradient through a gradient, which requests an additional backward pass through the function $f$ to compute Hessian-vector products \cite{finn2017model}. To simplify outer loop updates, we use a first-order approximation \cite{finn2017model} that eliminates the need for second-order derivatives. As the final outcome of the outer loop, we obtain the optimal parameters $\theta^*$. We then initialize our base model TE with these optimal parameters~$\theta^*$, resulting in the optimized base model $f_{\theta^*}$.  This optimized base model TE ($f_{\theta^*}$) now possesses enhanced adaptability, enabling it to quickly adapt to new, unseen abnormal trace classification tasks from any MSS.

\subsubsection{Meta-testing} In this phase, we use the optimized TE ($f_{\theta^*}$) to adapt to new, unseen abnormal trace classification tasks (meta-testing tasks) from any MSS. We configure each meta-testing task using the same settings as each meta-training task, denoting it as $T_{ts} = (S_{ts}, Q_{ts})$.  To adapt to a certain meta-testing task $T_{ts}$, we apply $f_{\theta^*}$ on its $S_{ts}$ for fine-tuning $f_{\theta^*}$'s parameters specifically for $T_{ts}$, resulting in adapted parameters ${\theta'_{ts}}$. After adaptation, we update $f_{\theta^*}$ to $f_{\theta'_{ts}}$, and use $f_{\theta'_{ts}}$ to classify abnormal traces in $Q_{ts}$ into fault categories for evaluating the performance of our framework TraFaultDia.

\section{Evaluation}



\subsection{Experimental design}
\label{sec:ex_design}
\subsubsection{Dataset} To train and evaluate our framework, we use trace data of TrainTicket and OnlineBoutique from DeepTraLog and Nezha. Detailed descriptions of these MSS and open datasets have been presented in Section \ref{sec:emprical_study}. As described in Section \ref{sec:emprical_study}, we establish our fault dataset (using DeepTraLog and Nezha), which includes abnormal traces from 30 fault categories for TrainTicket and 32 fault categories for OnlineBoutique. For our experiments, we further operate our fault dataset.  We divide 30 fault categories of Trainticket into 20 base and 10 novel fault categories, and 32 fault categories of OnlineBoutique into 22 base and 10 novel fault categories. That is, there are 20 and 22 base fault categories in TrainTicket and OnlineBoutique respectively, along with 10 novel fault categories for each system. To provide a consistent basis to compare experimental results, we standardized the number of novel fault categories for both systems. We design the composition of both base and novel fault categories for each system to include a random mix of fault categories from our dataset. This mix incorporates fault categories that comprise traces with varying numbers of spans and logs. It ensures the representation of fault categories with abnormal traces containing both more and fewer spans and logs within base and novel fault categories for each system. The full list of base and novel fault categories for each system is provided in our replication package.

\subsubsection{Training and evaluation} 
\label{sec:evaluation:training} Our training and evaluation process is designed based on our framework workflow (Section \ref{sec:ourFrameworkDesign}). We randomly selected 3960 unlabeled traces (3360 training/570 validation) from each MSS (TrainTicket/OnlineBoutique) to train AttenAE to construct trace representations for each MSS. We ensured that, in each training and validation set, normal traces largely outnumber anomalous ones to better reflect practical contexts. Besides, these traces do not overlap with those in our fault datasets.

To train and evaluate TEMAML, we use our fault dataset, which includes 20 and 22 base fault categories for TrainTicket and OnlineBoutique respectively, and 10 novel fault categories for each MSS. For each MSS, we randomly generate 4 meta-training tasks using its base fault categories, and 50 meta-testing tasks using its novel fault categories. Considering prior meta-learning studies~\cite{ye2021train,finn2017model} and the constraint of our fault dataset, we configure each meta-training task $T_{i}$ and each meta-testing task $T_{ts}$ in 5-way 5-shot and 5-way 10-shot setups: each $T_{i}$ and $T_{ts}$ is a distinct abnormal trace classification task with 5 fault categories, each category has 5/10 labeled trace instances as the support set ($T_{i}$, $T_{ts}$=5/10) for fine-tuning TEMAML to this task. For each $T_{i}$ and $T_{ts}$, we evaluate TEMAML on the query set consisting of 15 trace instances per fault category ($Q_{i}, Q_{ts}$=15). Specifically, for each MSS, we generate 4 meta-training tasks by performing 4 iterations of selecting a unique permutation of 5 fault categories from this MSS's base fault categories, and 50 meta-testing tasks by performing 50 iterations of selecting a unique permutation of 5 fault categories from this MSS's novel fault categories. The number of unique permutations for selecting 5 out of 10 novel fault categories is 252. We focus on 50 distinct meta-testing tasks, representing about 20\% of 252 unique permutations, to evaluate our framework, following the prior studies \cite{ye2021train,finn2017model}. This sampling approach allows for a focused comprehensive evaluation, ensuring various combinations of fault categories while keeping the data size within practical limits. For each MSS, TEMAML uses the optimized AttenAE’s encoder for this MSS to construct trace representations for abnormal traces in this MSS's meta-training and meta-testing tasks. With meta-training and meta-testing tasks for both MSS, we design four experiments E1-E4. E1 and E2 are within-system experiments for RQ1, while E3 and E4 are cross-system experiments for RQ2:  
\begin{itemize}
    \item \textbf {E1 (TrainTicket→TrainTicket).} We train TEMAML on TrainTicket's 4 meta-training tasks and evaluate it on TrainTicket's 50 meta-testing tasks. 
    \item \textbf{E2 (OnlineBoutique→OnlineBoutique).} We train TEMAML on OnlineBoutique's 4 meta-training tasks and evaluate it on OnlineBoutique's 50 meta-testing tasks. 
   \textbf{ \item E3 (OnlineBoutique→TrainTicket).} We train TEMAML on OnlineBoutique's 4 meta-training tasks and evaluate it on  TrainTicket's 50 meta-testing tasks. 
    \textbf{ \item E4 (TrainTicket→OnlineBoutique).} We train TEMAML on TrainTicket's 4 meta-training tasks and evaluate it on  OnlineBoutique's 50 meta-testing tasks. 
\end{itemize}





\subsubsection{Implementation details} E1-E4 are conducted on a Linux server with a 32-core CPU and an NVIDIA Ampere A100 GPU with 40 GB of memory, utilizing Python 3.10.6.  We train AttenAE and TEMAML using the AdamW optimizer. Further details regarding the implementation and hyperparameter settings of AttenAE and TEMAML are provided in our replication package \cite{TraFaultDia}.

\subsubsection{Baselines}
\label{sec:baselines}
Given the absence of directly comparable AIOps methods within this domain, we consider prior trace representation and classification methods to build our baselines. We combine baselines with ablation studies to systematically explore alternatives to evaluate the performance of our framework and the impact of each components within it:

\begin{table}[!ht]
\vspace{-7pt}
\centering
\small
\label{tab:baselines}
\begin{tabular}{l}
\hline
\textit{AttenAE alternative:} \\
Feed \textbf{only spans (OnlySpan)} into TEMAML for classification. \\

\textit{Multihead attention fusion alternatives:}\\
\textbf{Linear-based AE (LinearAE) fusion +TEMAML} \\
\textbf{Gated linear unit-based AE (GluAE) fusion +TEMAML } \\

\textit{Transformer encoder alternatives:} \\
   AttenAE+\textbf{LinearMAML} \\
   AttenAE+\textbf{RnnMAML} \\
   AttenAE+\textbf{LstmMAML}\\
   AttenAE+\textbf{CnnMAML}\\

\textit{TEMAML alternatives:} \\
    AttenAE+\textbf{TE-based Matching network (TEMatchNet)} \\
    AttenAE+\textbf{Prototypical network (ProtoNet)}   \\
    AttenAE+\textbf{Nearest neighbor (NNeighbor)}  \\
    AttenAE+\textbf{Decison tree (DTree)}\\
\hline
\end{tabular}

\vspace{-7pt}
\end{table}

AttenAE and multihead attention fusion alternatives incorporate different approaches than ours for constructing trace representations. The baseline OnlySpan follows the related work~\cite{nedelkoski2019anomaly2} to consider each trace as a sequence of spans and construct trace representations using only our identified span attributes. GluAE uses gated linear units to construct trace representations by fusing the same span and log attributes as our framework, modified from the modality fusion method for MSS in the study~\cite{lee2023eadro}. LinearAE is a simplified version of GluAE, utilizing linear projection to do fusion without the gating mechanism. 

We consider many TE alternatives, including the basic linear model, sequence models (RNN and LSTM), and CNN, as the alternative base models to perform multi-class classification on abnormal traces. We include sequence models here because prior studies \cite{kohyarnejadfard2019system,zhang2022putracead} have shown their effectiveness in performing binary classifications for normal and abnormal traces. CNN was considered as it has been used in related work~\cite{nedelkoski2019anomaly2} to do multi-class abnormal trace classification, as discussed in Section \ref{sec:sec_related_work}. The basic linear model serves as the simplest approach for evaluation.

TEMAML alternatives use other meta-learning methods (Prototypical network  \cite{snell2017prototypical}, matching network \cite{vinyals2016matching}) and traditional models (Nearest neighbor and Decision tree) for multi-class classification of abnormal traces. These alternatives completely replace TEMAML in our framework, exploring different efficient strategies to handle abnormal trace classification tasks. The fundamental machine learning models used in these alternatives struggle to effectively adapt to new, unseen abnormal trace classification tasks within-system or cross-system due to their simplicity and reliance on feature similarity for classification. Thus, we only evaluate these alternatives on E1 and E2 without transfer learning. For meta-testing tasks in E1/E2, we train each TEMAML alternative using their support sets and evaluate it on their query sets. This process mirrors the adaptation process in our framework—trace instances from each meta-testing task's support set are used for fine-tuning our framework before evaluation. The motivation for using these baselines is: if they can categorize abnormal traces as effectively and quickly by learning from labeled instances from those support sets, transfer learning may not be necessary for them.


We use the same neural representation method \cite{le2021log} as in our framework to handle textual span and log attributes in all baselines, although some of the previous studies referenced in our baselines utilized parsing and template approaches. The purpose is to ensure a fair comparison and demonstrate that the performance of our framework does not solely stem from the use of this neural representation method. This decision is driven by an empirical study \cite{le2021log}, which shows that neural representations are more effective at capturing the semantic meaning of textual log data.

\subsubsection{Evaluation metrics} 
To evaluate \textbf{effectiveness} of our TraFaultDia and baselines in fault categorization, we chose accuracy as the evaluation metric, a standard measure for multi-class classification tasks. Accuracy may not be the ideal metric in contexts where the distribution of classes is uneven. For instance, in anomaly detection-related binary classification tasks, normal traces often significantly outnumber abnormal ones, and a model could achieve high accuracy simply by predicting the most frequent class—normal traces. However, accuracy is suitable for our study as we perform multi-class classification tasks on abnormal traces in a 5-way meta-learning setup in each experiment. This setup ensures that consistently predicting a single class would yield a maximum accuracy of only 20\%. Using accuracy aligns with prior meta-learning studies \cite{finn2017model,ye2021train} on multi-class classification using accuracy as the only metric. 
To mitigate the risk of obtaining an inaccurate evaluation of the approach's effectiveness, in each experiment, we calculate the accuracy for each meta-testing task by conducting five trials and selecting the highest accuracy achieved. As evaluation results, we report the average accuracy along with the 95\% confidence interval (CI) computed across 50 meta-testing tasks in each experiment, as well as the range of task accuracies (minimum and maximum). This provides a comprehensive view of the approach's effectiveness and offers insights into its consistency and reliability in various scenarios. Besides, we perform t-tests~\cite{fisher1970statistical} to statistically compare the accuracy of our framework with effective baselines across 50 meta-testing tasks in each experiment.  We further quantify the differences using 
 Cohen's D value to assess the effect size. According to Cohen ~\cite{cohen2013statistical}, a value of 0.2 indicates a small effect size, 0.5 signifies a medium effect size, and 0.8 or higher denotes a large effect size. To measure the \textbf{efficiency} of our TraFaultDia and baselines in model training and evaluation, we calculate each's training time (spent on meta-training tasks) and testing time (spent on meta-testing tasks) in each experiment. Since AttenAE, a component of TraFaultDia, is trained prior to these tasks, we also calculate the training time of AttenAE on unlabeled traces in each MSS (as described in Section \ref{sec:evaluation:training}), as well as its testing time, i.e., the time taken by the optimized AttenAE to prepare trace representations for meta-testing tasks.
 
 


\subsection{Experiment results}
\subsubsection{Effectiveness.} 
\label{sec:effect_results}

Table \ref{tab:Trainticket_Result}-\ref{tab:Boutique_Results} shows our effectiveness evaluation results. 
Our TraFaultDia demonstrates high average accuracy on each experiment's 50 meta-testing tasks in both 5-shot and 10-shot setups, indicating its effectiveness in both within-system and across-system contexts. The four most effective baselines achieve an average accuracy of over 80\% with 95\% CI in at least one experimental setup: ``GluAE+TEMAML'', ``LinearAE+TEMAML'', and ``AttenAE+CnnMAML'', and ``AttenAE+NNeighbor''. Compared to these effective baselines, our TraFaultDia demonstrates better robustness by maintaining more consistent high accuracy across all experimental setups. Specifically, compared to our TraFaultDia, ``GluAE+TEMAML'' and ``LinearAE+TEMAML'' achieve similar high average accuracy in E1 and E3 setups, but are lower by around 2\%-8\% and 7\%-10\% in E2 and E4 setups respectively; ``AttenAE+CnnMAML'' achieve around 79.01-84.08\% average accuracy in E2 and E4 setups, but only obtain 49.04-69.47\% average accuracy in E1 and E3 setups. Though ``AttenAE+NNeighbor'' only achieves about 1\%-4\% lower average accuracy than our TraFaultDia in E1 and E2 setups concerning within-system adaptability, it has limited cross-system adaptability (i.e., transfer learning ability) as described in Section \ref{sec:baselines}.

\begin{table*}[!ht]
\centering
\scriptsize
\caption{Comparison of our TraFaultDia and baselines on TrainTicket’s 50 meta-testing tasks: \textit{Average accuracy with 95\% CI and min-max range}. E1 (TrainTicket→TrainTicket): train on TrainTicket's 4 meta-training tasks. E3 (OnlineBoutique→TrainTicket): train on OnlineBoutique's 4 meta-training tasks.}
\label{tab:Trainticket_Result}
\begin{tabular}{p{2.6cm} l l  l l}
\hline
Model & \multicolumn{2}{c}{E1 (TrainTicket→TrainTicket) } & \multicolumn{2}{c}{E3 (OnlineBoutique→TrainTicket)} \\
 \cmidrule(l){2-5}
& 5-shot & 10-shot & 5-shot & 10-shot \\
\hline
\textbf{Our TraFaultDia} & 92.91$\pm$2.10 (74.67-100.0) & 93.26$\pm$1.40 (76.00-100.0) & 86.35$\pm$2.00 (70.67-100.0) & 92.19$\pm$1.99 (74.67-100.0)\\
\textit{AttenAE alternative:}& & & & \\
\textbf{OnlySpan}+TEMAML & 80.64$\pm$2.84 (57.33-97.33) &  78.77$\pm$2.80 (60.00-97.33) & 79.25$\pm$2.89 (57.33-97.33) & 80.19$\pm$3.10 (56.00-97.33)\\
\multicolumn{2}{l}{\textit{Multihead attention fusion alternatives:}} & & & \\
\textbf{LinearAE}+TEMAML &  89.15$\pm$2.29  (73.33-100.0)&  90.59$\pm$2.43 (70.67-100.0) &    83.09$\pm$2.55 (62.67-97.33) & 90.61$\pm$2.01  (72.00-100.0) \\
\textbf{GluAE}+TEMAML & 92.21$\pm$1.73  (77.33-100.0) & 93.07$\pm$1.64 (77.33-100.0) & 85.07$\pm$2.38 (66.67-100.0) & 94.40$\pm$2.19 (72.00-100.0)  \\

\multicolumn{2}{l}{
\textit{Transformer encoder alternatives:}} & & & \\
 AttenAE+\textbf{LinearMAML} &  45.84$\pm$2.21 (25.33-61.33) &  45.36$\pm$2.16 (30.67-60.00)& 43.81$\pm$1.99 (28.00-58.67) &  43.87$\pm$1.93 (29.33-60.00) \\
 AttenAE+\textbf{RnnMAML}  & 49.65$\pm$2.09 (37.33-64.00) & 42.88$\pm$1.93 (24.00-58.67) & 48.45$\pm$1.75 (38.67-65.33) & 47.07$\pm$1.88 (34.67-58.67)\\
 AttenAE+\textbf{LstmMAML}  & 41.39$\pm$2.20 (21.33-56.00) & 42.67$\pm$1.91 (29.33-56.00)  & 40.32$\pm$1.68 (22.67-52.00) & 42.29$\pm$1.79 (25.33-56.00)  \\
 AttenAE+\textbf{CnnMAML}  & 57.06$\pm$2.85 (41.33-81.00) & 69.20$\pm$2.19 (56.00-88.00) & 49.04$\pm$1.80 (38.67-64.00) & 69.47$\pm$2.65 (48.00-89.33)\\ 

\textit{TEMAML alternatives:}& & & & \\
AttenAE+\textbf{TEMatchNet}  & 76.56$\pm$2.80 (49.33-93.33) & 76.05$\pm$2.37 (50.67-94.67) & — & —  \\
AttenAE+\textbf{ProtoNet} & 57.25$\pm$0.03 (40.00-74.67) & 59.68$\pm$0.03 (44.00-76.00) & — & —   \\
AttenAE+\textbf{NNeighbor} & 88.19$\pm$0.02 (74.67-98.67) & 92.56$\pm$0.02 (78.00-100.0) & — & —  \\
AttenAE+\textbf{DTree}  & 66.80$\pm$0.03 (46.67-88.00) & 77.09$\pm$0.03 (54.67-96.00) & — & —   \\

\hline
\end{tabular}
\vspace{-3.7pt}
\end{table*}

\begin{table*}[!ht]
\centering
\scriptsize
\caption{ Comparison of our TraFaultDia and baselines on OnlineBoutique's 50 meta-testing tasks: \textit{Average accuracy with 95\% CI and min-max range}. E2 (OnlineBoutique→OnlineBoutique): train on OnlineBoutique's 4 meta-training tasks. E4 (TrainTicket→OnlineBoutique): train on TrainTicket's 4 meta-training tasks.}
\label{tab:Boutique_Results}
\begin{tabular}{p{2.6cm} l l l l}
\hline
Model & \multicolumn{2}{c}{E2 (OnlineBoutique→OnlineBoutique) } & \multicolumn{2}{c}{E4 (TrainTicket→OnlineBoutique)} \\
 \cmidrule(l){2-5}
& 5-shot & 10-shot & 5-shot & 10-shot \\
\hline

\textbf{Our TraFaultDia}  & 82.50$\pm$2.35 (65.33-98.67) & 85.20$\pm$2.33 (66.67-98.67) & 82.37$\pm$2.07 (64.00-97.33) & 84.77$\pm$2.28 (68.00-98.67) \\ 

\textit{AttenAE alternative:}& & & & \\

\textbf{OnlySpan}+TEMAML & 72.83$\pm$2.40 (57.33-88.00) & 73.15$\pm$2.81 (46.67-92.00) &  71.81$\pm$2.25 (56.00-85.33) &  73.60$\pm$2.21 (57.33-85.33)\\
\multicolumn{2}{l}{\textit{Multihead attention fusion alternatives:}} & & & \\
\textbf{LinearAE}+TEMAML & 76.15$\pm$2.59 (60.00-95.00) & 78.21$\pm$2.50 (64.00-96.00) &  75.81$\pm$2.45 (52.00-89.33) &  74.32$\pm$2.44 (53.33-88.00)\\
\textbf{GluAE}+TEMAML  & 80.61$\pm$2.96 (58.67-98.67) & 77.49$\pm$2.67 (48.00-94.67) & 74.96$\pm$2.76 (54.67-94.67) & 77.57$\pm$2.70 (56.00-94.67)\\

\multicolumn{2}{l}{\textit{Transformer encoder alternatives:}} & & & \\
 AttenAE+\textbf{LinearMAML} & 42.59$\pm$3.63 (20.00-77.33)  &  40.75$\pm$3.53 (21.33-68.00)& 47.01$\pm$3.59 (25.33-74.67) & 44.35$\pm$4.10 (20.00-89.33)  \\
 AttenAE+\textbf{RnnMAML}  &  72.59$\pm$2.50 (54.67-94.67) & 64.75$\pm$2.53 (46.67-80.00) & 72.58$\pm$2.58 (54.70-94.70) & 71.01$\pm$2.80 (56.00-89.33) \\
 AttenAE+\textbf{LstmMAML}  & 54.80$\pm$2.11 (40.00-70.67) & 55.97$\pm$2.25 (41.33-69.33) & 56.19$\pm$1.92 (38.67-72.00) & 59.71$\pm$2.05 (42.67-77.33) \\ 
AttenAE+\textbf{CnnMAML}  & 80.10$\pm$2.16 (60.00-94.67) & 83.07$\pm$3.29 (68.00-97.33)& 79.01$\pm$2.63 (56.00-97.33) & 84.08$\pm$2.76 (65.33-100.0) \\ 

\textit{TEMAML alternatives:}& & & & \\
AttenAE+\textbf{TEMatchNet} & 76.29$\pm$3.00 (54.67-96.00) & 73.11$\pm$2.94 (50.67-94.67) & — & —  \\
AttenAE+\textbf{ProtoNet} & 74.51$\pm$0.03 (53.33-92.00) & 76.59$\pm$0.04 (58.66-94.67)  & — & —   \\
AttenAE+\textbf{NNeighbor} & 80.96$\pm$0.03 (64.00-98.67) & 84.75$\pm$0.03 (62.80-98.67) & — & — \\

AttenAE+\textbf{DTree}  &  66.99$\pm$0.02 (54.67-80.00) & 73.79$\pm$0.03 (58.67-82.67)  & — & —  \\

\hline
\end{tabular}
\vspace{-3.7pt}
\end{table*}

\begin{table}[!ht]
\centering
\small
\captionsetup{justification=centering}
\caption{T-test results of accuracy on 50 meta-testing tasks in each experiment, \\ comparing our TraFaultDia to  effective baselines in our experiments.}
\label{tab:t_tests}
\begin{tabular}{l l lllll}
\hline
\textbf{Our TraFaultDia v.s.} & E1 & 5-shot & 10-shot & E3 & 5-shot & 10-shot\\ \hline 
 
GluAE+TEMAML &  &   0.0719 & 0.5347 &  &  0.0045*** & 7.74e-07*** \\
LinearAE+TEMAML &  &  1.62e-13*** 
& 1.14e-09*** 
 &  & 1.87e-10*** & 0.0005*** \\
AttenAE+NNeighbor & 
 &  7.10e-29***
 & 0.0006*** 
  &  & — & —  \\

\hline

 \textbf{Our TraFaultDia v.s.} & E2 &  5-shot & 10-shot & E4  & 5-shot & 10-shot\\
 \midrule
GluAE+TEMAML &  &  0.0006*** &6.95e-28*** &  &  1.70e-27*** & 9.15e-27*** \\
AttenAE+CnnMAML&  &  6.65e-07*** & 0.0003*** &  &  2.01e-10*** & 0.176 \\
AttenAE+NNeighbor &  &  1.11e-05*** & 0.1752  &   & —  & —   \\
\hline
\end{tabular}
\begin{center}
\footnotesize ***p < 0.001; **p < 0.01; *p < 0.05
\end{center}
\vspace{-12pt}
\end{table}

\begin{table}[!ht]
\centering
\small
\caption{Effect size: Cohen's d values comparing our TraFaultDia to effective baselines.}
\label{tab:effectSize}
\begin{tabular}{lllllll}
\hline
\textbf{Our TraFaultDia v.s.} & E1 & 5-shot & 10-shot & E3 & 5-shot & 10-shot\\ \hline
 
GluAE+TEMAML & & 0.36 & 0.13 & & 0.58 & -1.06  \\
LinearAE+TEMAML & & 1.71 & 1.35 & & 1.42 & 0.79 \\
AttenAE+NNeighbor & & 3.18  & 0.71  & & — & —  \\

\hline

 \textbf{Our TraFaultDia v.s.} & E2& 5-shot & 10-shot & E4 & 5-shot & 10-shot\\
 \hline
GluAE+TEMAML & &0.71  & 3.07 & & 3.04 & 2.88  \\
AttenAE+CnnMAML & &1.06 & 0.75 & & 1.42 & 0.27 \\
AttenAE+NNeighbor & &0.93 & 0.27  & & —  & —   \\\hline
\end{tabular}
\vspace{-8pt}
\end{table}

Table \ref{tab:t_tests} presents t-test results of accuracy on each experiment's 50 meta-testing tasks, comparing our TraFaultDia to the
three most effective baselines in each experiment. 
Except for ``GluAE+TEMAML'' in E1 setups, ``AttenAE+NNeighbor'' in the E2 10-shot setup, and ``AttenAE+ CnnMAML'' in the E4 10-shot setup, our TraFaultDia demonstrates significant differences from baselines (p < 0.001) in all other E1-E4 setups. Table \ref{tab:effectSize} compares Cohen's d values of the accuracy of our TraFaultDia to these baselines across 50 meta-testing tasks in each experiment. Our TraFaultDia demonstrates large positive effect sizes (Cohen's~d~>~0.8) against ``GluAE+TEMAML'' primarily in E2 and E4, and against ``LinearAE+TEMAML'' and ``AttenAE+NNeighbor'' in 5-shot setups across all experiments. The above t-test results and effect size measurements provide strong evidence for the effectiveness of our framework over these baselines in most experimental steups.

\textit{Exploration results: }For OnlineBoutique's meta-testing tasks in E2 and E4 (Table \ref{tab:Boutique_Results}), the lowest accuracies (48\%-68\%) for both our TraFaultDia and the three most effective baselines occur in classifying abnormal traces caused by performance issues ``CPU contention'', ``CPU consumption'', and ``network delay'' on the same pods. These instances significantly impact the overall average accuracy in E2 and E4. Incorporating performance metrics such as CPU, memory, and network traffic into the construction of trace representations could improve the categorization accuracy of such faults by both our TraFaultDia and baseline approaches.


\subsubsection{Efficiency}


Table \ref{tab:our_framework_training_time}-\ref{tab:evaluation_result_adaptability} compares the training and testing times of our TraFaultDia against the
three most effective baselines in each experiment. As shown in Table \ref{tab:our_framework_training_time}, compared to these baselines, our TraFaultDia takes about 6-73 seconds less in some E3 setups, while it takes approximately 4-29 seconds more training time compared in E1, E2, E4. As shown in Table \ref{tab:evaluation_result_adaptability}, MAML-related approaches, including our TraFaultDia, take around 4-11 times less testing times than ``AttenAE+NNeighbor'' in E1-E4. ``AttenAE+NNeighbor'' uses NNeighbor to compare every new trace to all existing labeled instances \cite{Cover1967nneighbor}, resulting in increased testing time as the dataset expands. As the number of abnormal trace classification tasks increases in real-world operational contexts of MSS, the time required for ``AttenAE+NNeighbor'' to adapt to each task can accumulate, leading to significantly longer adaptation times on new abnormal trace categorization tasks than MAML-related methods.  



\begin{table}[!ht]
\centering
\small
\caption{ Training time$^a$ (in seconds) of our TraFaultDia and effective baselines.}
\label{tab:our_framework_training_time}
\begin{tabular}{l | p{0.35cm}  r  r | p{0.35cm} r r}
\toprule
 & E1 & 5-shot & 10-shot  & E3 & 5-shot & 10-shot\\
 \midrule
\textbf{Our TraFaultDia} & & 36.3394 & 25.8387 &  & 46.0045 &  62.7578 \\
GluAE+TEMAML & & 19.9185& 20.6976 & & 42.3725 & 78.0125 \\
LinearAE+TEMAML & &  17.7576 & 45.1150 & & 119.6440 & 30.5474  \\
AttenAE+NNeighbor*  & & — & — & & — & — \\

\midrule

  & E2 & 5-shot & 10-shot &  E4 & 5-shot & 10-shot\\
 \midrule
\textbf{Our TraFaultDia} & & 40.9426 & 65.3120 & & 32.4136 & 53.3256 \\
GluAE +TEMAML & &20.1432 & 58.8608 & & 21.6602 & 47.0818    \\
CnnMAML & & 23.9832 & 36.7725 & & 15.6358 & 23.6530 \\
AttenAE+NNeighbor*  & & — & — & & — & —     \\
\bottomrule
\end{tabular}
\tiny
\begin{flushleft} $^a$Training time refers to the time each method used to train on meta-training tasks in each experiment. It excludes the time spent on constructing trace representations to avoid variations caused by different trace representation approaches.  \newline *As described in Section \ref{sec:baselines}, for each meta-testing task in our experiments, we train ``AttenAE+NNeighbor'' on this task's support instances, treating this as the adaptation process to it, without using meta-training tasks. Thus, there is no training time on meta-training tasks for ``AttenAE+NNeighbor''.
\end{flushleft}

\vspace{-5pt}
\end{table}

\begin{table}[!ht]
\centering
\small
\caption{Testing time$^b$ per meta-testing task (in seconds) of our TraFaultDia and effective baseline.}
\label{tab:evaluation_result_adaptability}
\begin{tabular}{l | p{0.35cm} l  l | p{0.35cm} l l}
\toprule
 & E1 & 5-shot & 10-shot & E3 & 5-shot & 10-shot\\
 \midrule
\textbf{Our TraFaultDia} & & 0.0460 & 0.0680 & & 0.0651 & 0.0953   \\
GluAE+TEMAML & & 0.0640 & 0.0942 && 0.0651 & 0.0954  \\
LinearAE+TEMAML & & 0.0748 & 0.0946 && 0.0640 & 0.0944 \\
AttenAE+NNeighbor & & 0.1860  & 0.1890  &&  — & —  \\

\midrule

 & E2& 5-shot & 10-shot & E4 & 5-shot & 10-shot\\
 \midrule
\textbf{Our TraFaultDia} & & 0.0848 & 0.0977 & & 0.0875 & 0.0974   \\
GluAE +TEMAML & &0.0801  & 0.0978 &  &0.0731 & 0.0969  \\
CnnMAML & & 0.0470 & 0.0472 & & 0.0417 &0.0490 \\
AttenAE+NNeighbor & & 0.5140 & 0.5307   && —  & —   \\
\bottomrule
\end{tabular}
\begin{flushleft} \scriptsize $^b$Testing time refers to the time each method is evaluated on query sets $Q_{ts}$ of 50 meta-testing tasks in each experiment.
\end{flushleft}
\end{table}

 
 Table \ref{tab:training_effi} compares the training time of AttenAE in our TraFaultDia  with baseline trace representatin approaches for each MSS. Table \ref{tab:trace_cons_effi} presents the average time AttenAE and these baseline approaches take to construct trace representations per meta-testing task  (covering both 5-shot and 10-shot setups) for each MSS. As shown in Table \ref{tab:training_effi}-\ref{tab:trace_cons_effi}, our AttenAE, GluAE, LinearAE require a similar amount of training time and trace construction time. OnlySpan takes around half the training and trace construction time compared to other approaches because it completely omits logs in trace representation construction; however, the OnlySpan based baseline ``OnlySpan+TEMAML'' achieves approximately 10\% lower average accuracy in E1-E4 setups compared to others using different approaches, referring to Table~\ref{tab:Trainticket_Result}-\ref{tab:Boutique_Results}.
 

\begin{table}[!ht]
\centering
\begin{minipage}[t]{0.42\textwidth}
    \centering
    \small
    \captionsetup{justification=centering}
    \caption{Training time (in minutes) of trace construction approaches for each MSS}
    \label{tab:training_effi}
    \begin{tabular}{lll}
    \toprule
    & TrainTicket & OnlineBouque \\ \hline
    \textbf{Our AttenAE} & 22.4112 & 18.7502\\
    GluAE & 20.1517 & 18.6413 \\
    LinearAE & 20.5443 & 15.5549\\
    OnlySpan & 13.4135 & 10.3123 \\
    \bottomrule
    \end{tabular}
\end{minipage}
\hfill
\begin{minipage}[t]{0.52\textwidth}
    \centering
    \small
    \captionsetup{justification=centering}
    \caption{Average trace construction time (in seconds) per meta-testing task}
    \label{tab:trace_cons_effi}
    \begin{tabular}{lllll}
    \toprule
     & \multicolumn{2}{c}{Trainticket} & \multicolumn{2}{c}{OnlineBoutique} \\
     \cmidrule(l){2-5}
     & 5-shot & 10-shot & 5-shot & 10-shot\\\hline
    \textbf{Our AttenAE}  & 1.8436 & 2.1524  & 4.1426 & 5.0189 \\
    GluAE & 1.8395  & 2.1541 & 4.3121 & 5.1921  \\
    LinearAE & 1.8286  & 2.1486 & 4.1223 & 5.2066 \\
    OnlySpan & 1.1671 & 1.1818 & 1.9065 & 2.1822  \\
    \bottomrule
    \end{tabular}
\end{minipage}%
\vspace{-5pt}
\end{table}

\subsubsection{Answers to our research questions:} Our evaluation results indicate that, our TraFaultDia, once adequately trained, can quickly adapt to new abnormal trace classification tasks from any MSS using just a few labeled traces; it demonstrates robust effectiveness (i.e., high accuracy) with fast adaptation times compared to baselines. Thus, we conclude that, TraFaultDia demonstrates effective and fast within-system and cross-system adaptability for classifying abnormal traces into precise fault categories for MSS.

\subsection{Ablation study} As compared to the different alternative baselines shown in Table \ref{tab:Trainticket_Result}-\ref{tab:Boutique_Results}, each component of our TraFaultDia contributes to its overall effectiveness, with AttenAE is the most significant contributor. The impact of AttenAE is evident from these tables, where our TraFaultDia consistently outperforms  the AttenAE alternative ``OnlySpan+TEMAML" by approximately 10\% in E1-E4 setups. Additionally, we conduct an ablation study on the number of meta-learning tasks to train TEMAML. Our current study uses 4 meta-training tasks in our experiments.  We tested 2-4 meta-training tasks and found that using 4 meta-training tasks yielded the best average accuracy across 50 meta-testing tasks in each experiment, see Table \ref{tab:Ablation_meta_traning_tasks}. This result aligns with the nature of MAML as a multi-task learning algorithm: including more varied meta-training tasks can enhance the algorithm's robustness, improving its ability to adapt to new, unseen tasks across different contexts.


\begin{table}[ht]
\centering
\small
\caption{Ablation study on the number of meta-training tasks in each experiment.}
\label{tab:Ablation_meta_traning_tasks}
\begin{tabular}{p{2.2cm} p{1cm} p{1.2cm} p{1cm} p{1.1cm} p{1cm} p{1.1cm} p{1cm} p{1.1cm}}
\hline
  & \multicolumn{2}{c}{E1} & \multicolumn{2}{c}{E2} & \multicolumn{2}{c}{E3} & \multicolumn{2}{c}{E4}  \\
 \cmidrule(l){2-9}
 Number of tasks &  5-shot & 10-shot & 5-shot & 10-shot & 5-shot & 10-shot & 5-shot & 10-shot\\
 \midrule
\textbf{Current: 4} & 92.91 & 93.26 & 82.50 & 85.20 & 86.35 & 92.19 & 82.37 & 84.77 \\
3 & 87.67  & 90.00 & 80.53 & 82.13 & 85.03 & 89.07 & 81.92 & 82.67\\
2 & 88.17 & 89.33 & 78.67 & 78.10 & 83.11 & 85.78 & 74.67 & 77.33 \\
\hline
\end{tabular}
\end{table}


\section{Threats to validity}

External threats may arise from the limitation of our fault dataset sourced from open datasets (DeepTraLog and Nezha) of benchmark systems. First, using datasets from real-world systems would increase the applicability and relevance of our study, ensuring that the findings are more reflective of actual operational environments. Due to our limited resources, we are unable to access comprehensive real-world datasets. Second, we intended to include more trace-related metrics in constructing representations for abnormal traces. Including trace-related performance metrics could potentially help distinguish performance-related anomalies. This
may address the issue referred to our exploration results in Section \ref{sec:effect_results}. Nezha \cite{Nezha} provides some performance metrics, e.g., CPU, memory usage, and network traffic, related to traces over time. We did not find other MSS open datasets with such rich modalities. Third, we aimed to expand the evaluation of our framework by including additional datasets that contain abnormal traces from a broader range of fault types, as well as datasets from various MSS. Despite our efforts, we were unable to find such datasets. Thus, our study used abnormal traces from 30 and 32 fault categories in TrainTicket and OnlineBoutique, respectively. This limitation led us to adopt a 5-way setup in order to ensure an adequate distribution of meta-training and meta-testing tasks in each experiment. However, prior studies have demonstrated that MAML is highly effective for complex multi-class classification tasks (e.g., 20-way and 50-way classifications)  across a variety of contexts \cite{jung2022few,jamal2019task,yu2020transmatch}. The above three external threats pose the possibility that the application of our framework for classifying abnormal traces for MSS might not fully utilize its generalization capabilities. Addressing these threats may improve our framework, thereby enhancing the precision and robustness of trace-level RCA in MSS. To address these threats, we are in the process of selecting and deploying real-world industrial MSS \cite{amoroso2024dataset} and generating new datasets.

\section{Conclusion}
This paper proposes a novel framework, TraFaultDia, which automatically classifies abnormal traces into specific fault categories for MSS. TraFaultDia has two main
components: (1) AttenAE for unsupervisedly constructing unified, compressed trace representations, which
enables (2) TEMAML to perform effective few-shot categorization of abnormal traces from MSS. The proposed framework is evaluated on representative benchmark MSS with open datasets. The evaluation results show that TraFaultDia achieves effective and rapid within-system and cross-system adaptability. Future work will focus on further improving generalizability, scalability, and interpretability of TraFaultDia. We plan to test its performance on real-world industrial MSS.

\section*{Acknowledgment}
This work is funded by the Research Council of Finland (Decision No. 349487) under the academic project MuFAno, and Ulla Tuominen Foundation. The authors acknowledge CSC-IT
Center for Science, Finland, for providing computing resources. The authors would like to thank Nauman bin Ali from the Blekinge Institute of Technology (Sweden) for reading
the paper and providing valuable feedback.

\bibliographystyle{ACM-Reference-Format}
\bibliography{own}

\end{document}